\documentclass{article}
\usepackage[utf8]{inputenc}
\usepackage{multirow}
\usepackage{makecell}
\usepackage{amsmath}
\usepackage{booktabs}
\usepackage{graphicx}
\usepackage{xcolor}
\usepackage{adjustbox}
\usepackage{amsmath,amsfonts,graphicx}
\usepackage{mathtools}
\usepackage{amssymb,bbm}
\usepackage{romanbar}
\usepackage{multirow}
\usepackage{hyperref}
\title{A Robust Statistical Analysis of the Role of Hydropower on the System Electricity Price and Price Volatility}
\author
{Olukunle O. Owolabi $^{1\ast}$, Kathryn Lawson$^{2,3}$,
\\Sanhita Sengupta$^{4}$, Yingsi Huang$^{5}$, Lan Wang$^{5}$, \\Chaopeng Shen$^{2}$, Mila Getmansky Sherman$^{6}$,
\\Deborah A. Sunter$^{1,7\ast}$ \\
\normalsize{$^{1}$Department of Mechanical Engineering, Tufts University, USA}\\
\normalsize{$^{2}$Department of Civil and Environmental Engineering, The Pennsylvania State University, USA}\\
\normalsize{$^{3}$ HydroSapient, Inc., USA}\\
\normalsize{$^{4}$School of Statistics, University of Minnesota, USA}\\
\normalsize{$^{5}$Department of Management Science, University of Miami, USA}\\
\normalsize{$^{6}$Isenberg School of Management, UMASS Amherst, USA}\\
\normalsize{$^{7}$Department of Civil and Environmental Engineering, Tufts University, USA}\\
\normalsize{$^\ast$To whom correspondence should be addressed, E-mail:  olukunle.oladipo@gmail.com}\\
or\\
\normalsize{deborah.sunter@tufts.edu}
}

\date{}
\begin{document}
\maketitle

\section*{Abstract}
Hydroelectric power (hydropower) is unique in that it can function as both a conventional source of electricity and as backup storage (pumped hydroelectric storage) for providing energy in times of high demand on the grid \cite{Rehman2015}. This study examines the impact of hydropower on system electricity price and price volatility in the region served by the New England Independent System Operator (ISONE) from 2014 - 2020 \cite{ISONE2021}. We perform a robust holistic analysis of the mean and quantile effects, as well as the marginal contributing effects of hydropower in the presence of solar and wind resources. First, the price data is adjusted for deterministic temporal trends, correcting for  seasonal, weekend, and diurnal effects that may obscure actual representative trends in the data. 
Using multiple linear regression and quantile regression, we observe that hydropower contributes to a reduction in the system electricity price and
price volatility. While hydropower has a weak impact on decreasing price and volatility at the mean, it has greater impact at extreme quantiles ($> $ 70th percentile). At these higher percentiles, we find that hydropower provides a stabilizing effect on price volatility in the presence of volatile resources such as wind. We conclude with a discussion of the observed relationship between hydropower and system electricity price and volatility.

\section{Introduction}
Satisfying electricity demand while reducing environmental impact of energy generation is critical in the growing ambition to combat climate change \cite{Buonocore2019, frass2009renewable} and has contributed to an increase in renewable energy deployment \cite{useiarenew2021}. Hydroelectric power (hydropower) is a large contributor to the renewable energy portfolio. According the the US Energy and Information Administration, 7.3\% of utility scale electricity in the United States is produced from hydropower, representing a substantial contribution to the US renewable electricity mix (all other forms of renewable generation account for 12.5\% with wind and solar representing 8.4\% and 2.3\%, respectively) \cite{EIA2021}.

A unique characteristic of hydropower is that it provides both conventional electricity generation and long-duration storage \cite{Rehman2015}. This complementary ability means that it can serve as a back up for variable renewable energy (VRE) sources (such as wind and solar) \cite{Foley2015} and provide some adaptive response to meet electricity demands. Despite some environmental drawbacks of hydropower, its added value as a storage system, which is even more necessary in the current global energy transition, has ignited renewed interest in the technology leading to rehabilitation of old hydropower systems and development of new systems as an economically and technically viable option to support volatile solar and wind resources \cite{Ardizzon2014}. Several studies have explored the feasibility of such hydro-wind \cite{Kaldellis2001, Bakos2002,Kapsali2010,Kapsali2012}, hydro-solar \cite{Glasnovic2009,Margeta2010,Margeta2012,Zhao2012,Javanbakht2013,Ma2015} and hydro-wind-solar \cite{Zhang2019} systems.

Hydropower may promote electricity price stability. Unlike coal and natural gas, it is not impacted by fossil fuel market fluctuations \cite{Ardizzon2014}. While there are seasonal variations due to the availability of water, large reservoirs can mitigate this risk \cite{Wen2022}. Hydropower provides more stable and predictable generation when compared to VREs. Some studies demonstrate that wind can contribute to an increase in electricity price volatility \cite{Woo2011,Ketterer2014}. 
In contrast, studies on the impact of hydropower on price volatility showed that dispatchable hydropower may contribute to a reduction in electricity price and price volatility \cite{Suomalainen2015,Pereira2017}.
 The extent of these relationships are fairly dynamic and may vary regionally depending on factors such as the amount of hydropower present, the composition of the market's energy portfolio, the market and incentive structure, and seasonal and temporal patterns of electricity demand.

 A novel contribution of this study is the analysis of quantile effects between hydropower and price and price volatility since previous studies have predominantly focused on quantifying mean effects \cite{Pereira2017,Somani2021,Wen2022}. The quantification of quantile effects gives us insight into the relationship at extreme electricity prices given the skewed nature of its distribution (Table \ref{tab:stat}). Additionally, this study provides an adjustment model that corrects for deterministic temporal trends (seasonal, hourly, or weekend) in the electricity price which, if uncorrected, may otherwise lead to erroneous conclusions. These contributions are important for creating a holistic understanding of the role of hydropower in system electricity price and price volatility, which could inform policy and decision-making for new and existing hydropower projects to meet ambitious renewable energy targets.      \par

This study also explores the mean and quantile relationship of hydropower  
on the system electricity price and price volatility in the presence of other forms of renewable energy like wind and solar. The study leverages robust statistical methods of Multiple Linear Regression (MLR) and Quantile Regression (QR) for a case study of the region served by the New England Independent System Operator (ISONE) from 2014 - 2020 \cite{ISONE2021}. An adjustment model is first used to exclude deterministic temporal effects that may obfuscate actual trends in the data before exploring both the mean and quantile relationships of hydropower, solar power, and wind power with system electricity price and price volatility. The study provides a robust holistic evaluation of the impact of hydropower on the wholesale electricity market that may inform the design of energy portfolios, policy incentives, and plans for new and existing hydropower plants.

\section{Study Area and Data}
The case study is based on ISONE, which is an independent non-profit company that coordinates grid operation, market administration, and power system planning for the New England states of Vermont, Connecticut, Rhode Island, Massachusetts, New Hampshire, and most of Maine \cite{ISONEabout2021}. 
ISONE has undergone substantial changes in terms of the diversity of its generation mix. In particular, there has been a significant addition of renewables to the grid in the past few years. Renewables (predominantly wind and solar resources) account for 9\%, while hydropower accounts for an additional 7\% of the electric energy generation in ISONE \cite{ISONEresource2021}. Though there is controversy on the classification of hydropower as a clean renewable source of electricity, it still represents a stable source of power for the region, and an opportunity to serve as a low cost method of energy storage. Roughly 2,000 MW of large-scale hydroelectric energy storage is available in ISONE \cite{isoneoutlook2020}. This offers the opportunity to balance electricity supply and demand, and reduce the volatility associated with the intermittency of VRE. 

The data used in the study is based on the real-time electricity price and generation data from ISONE from 2014-2020. 
ISONE gives data on the system price of electricity and the amount of electricity that was generated from various energy sources (e.g., hydro, solar, wind, natural gas, coal, nuclear). The ISONE data was obtained from the ISONE web services API v.1.1 \cite{ISONE2021} that gives a range of public market and energy data. ISONE reports information on the price paid at different nodes. Due to transmission and congestion costs, the price at two different nodes can be different. To ensure that the results are interpretable and comparable,
we remove transmission and congestion costs to compute a system price. The system price reflects the cost of generating one megawatt-hour of electricity. The system price is also known as the Marginal Energy Cost, MEC (see Eqn. \ref{eq. LMP}). 
We collected historical data on the following information:
\begin{enumerate}
\item Electricity prices: The hourly price data was obtained for the hub location in hourly resolution alongside the losses and congestion cost. The data contains information on the Locational Marginal Price, Congestion Price, and Losses. The system price or MEC, is then obtained as \cite{isonelmp}: 
\begin{equation} \label{eq. LMP}
    MEC = LMP - MCC - MLC 
\end{equation} 
where:  $LMP$ = Locational Marginal Price\\
 $MEC$ = Marginal Energy Cost\\
    $MCC$  = Marginal Congestion Cost\\
    $MLC$ = Marginal Loss Cost\\
\item Generation mix: This includes electricity generation, separated by energy source, used to satisfy demand. The generation data was obtained for the entire system in resolutions ranging from 5-15 minutes and then aggregated to hourly resolution. 
\end{enumerate}


\section{Methods}

\subsection{Adjustment Model: Seasonal, Diurnal and Weekend Effects}
 Electricity price is driven by changes in the characteristics of consumer demand. 
For example, energy prices are typically higher in the day (when consumers are awake) than at night, and higher in the summer (when consumers use air conditioning) than in the spring.
In order to properly analyse the impact of hydropower on energy prices, 
we need to take these patterns into account.
To do so, we use a linear model with categorical variables to estimate these temporal effects. 
Categorical variables for hour of day, season, and weekend were included in the model. The adjustment model allows for interaction effects between the hour of day and season, 
so that each season has its own diurnal pattern. 
More formally, for the hourly data, the system electricity price at time $t$ is given by: 
\begin{equation}\label{eq:hourly_lm}
    \text{Price}(t) = 
\beta_1 \text{hour}(t) + \beta_2 \text{season}(t) + \beta_3 \text{hour}(t)\times \text{season}(t) + \beta_{4} \text{weekend}(t) + \epsilon_t 
\end{equation}
where hour$(t)$, season$(t)$ give the hour of the day and season at time $t$ respectively,
weekend$(t)$ is a binary variable indicating whether time $t$ occurs on a weekend, and 
$\epsilon_t$ is a residual term.
We subtract the values of the predicted effects from the system electricity price to obtain the \textbf{detrended system electricity price}. This detrended system price was used throughout this study. 
Also the price volatility calculations in this study is based on the detrended system price. 

\begin{table}[h!]
\centering
\begin{tabular}{c c c c c c c}
	\hline
	~ & \makecell{Hydro\\(MWh)} & \makecell{Solar \\(MWh)} & \makecell{ Wind\\ (MWh)} & \makecell{Detrended Price \\(\$)} & \makecell{{Detrended Price} \\{Volatility (\$)}} \\
	\hline
	~~ min & 185.14 & 0.00 & 0.00 & -156.86 & 0.00 \\
	~~ max & 2606.33 & 218.85 & 1148.80 & 2446.71 & 1140.54\\
	~~ median & 826.68 & 0.25 & 290.80 & 25.85 & 2.57\\
	~~ mean  & 864.62  & 12.22 & 339.63 & 34.43  & 5.48\\
	~~ std & 415.05 & 26.81 & 244.10 & 36.09 & 15.16\\
    \hline
\end{tabular}
\caption{\textbf{Descriptive Statistics Summary.} This shows the minimum, maximum, median, and mean statistics for hydro, solar, and wind generation (in MWh), and detrended system price and price volatility (in \$) for ISONE between 2014 - 2020. (Note: std = standard deviation)}
\label{tab:stat}
\end{table}

\subsection{Price Volatility and Energy Penetration Calculations}\label{calc}
We calculated the volatility of detrended system electricity price as a function of time (temporal price volatility), using the Exponential Weighted Moving Standard Deviation (EWMSD). 
Detrended price volatility is the instantaneous standard deviation of the hourly detrended electricity price. The exponentially weighted moving standard deviation improves on simple standard deviation by assigning weights to the periodic price. Specifically, EWMSD helps to account for right positive skew nature of the price distribution.

\par We also calculated the penetration of each of energy source (hydropower, wind and solar) as a percentage of the amount of energy generated that hour by each source, respectively, relative to the total energy generated that hour.

\subsection{Mean Effects of Hydropower Penetration on Detrended System Electricity Price and Detrended Price Volatility}
Multiple Linear Regression (MLR) was used to analyse the mean relationship between hydropower penetration (as calculated in section \ref{calc}) and detrended electricity price and detrended price volatility. Four different models were built, with VRE penetrations as independent variables: (1) hydropower, (2) hydropower + solar, (3) hydropower + wind, (4) hydropower + wind + solar. These models provide insight on the average relationship between hydropower and detrended price and detrended price volatility, as well as the contributing effects in the presence of these other sources of renewable electricity. By analysing the model coefficients, standard error, and statistical significance (p-value) we were able to describe the extent of the mean relationship between hydropower, price, and price volatility. To ascertain that multicollinearity assumptions \cite{Yoo2014} for MLR were met, specifically that little or no multicollinearity exist in the predictor variables used in the models, a correlation plot was used to observe the extent of collinearity between the variables (see Figure \ref{fig:correlation plot}). 

\subsection{Quantile Effects of Hydropower Penetration on Detrended System Electricity Price and Detrended Price Volatility}
We used Quantile Regression (QR) to understand the quantile effect of hydropower on the system electricity price and price volatility as well as joint quantile effects in the presence of other renewables, specifically solar and wind. QR is useful in modelling the conditional relationship between a predictor and a response variable at the median or conditional points in the distribution \cite{Bernard2015}. QR is a robust statistical method which is less affected by outliers in the data set compared to its least-squares regression counterpart \cite{lane20212}. \par
For a given a response variable $Y$ and a predictor $X$, QR is used to estimate the relationship between $X$ and the
conditional quantile of $Y$.  Mathematically, the $\tau$th ($0<\tau<1$) conditional quantile of $Y$ given 
$X=x$ is defined as $q_{Y}(\tau|x) = \inf\{t:f_{Y|x}(t)\geq\tau\}$, where $f_{Y|x}$ is the conditional cumulative
distribution function of $Y$ given $X=x$. 
  Interpretation of the conditional quantile is straightforward.  For example, given the predictor value $X=x$ and $\tau=0.5$, 50\% of observations of $Y$
with associated $x$ fall below the conditional median, $Q_{Y|X}(0.5)$. 

\par With this, we were able to describe not just the average relationship (like in MLR), 
but also the effects at the extremes of the distribution. This is particularly useful for detrended price and detrended volatility since they have a skewed distribution as seen in Table \ref{tab:stat}. Effects at the extremes may differ from those at the mean; QR can reveal patterns that occur at those extremes. Like the mean effects analysis, four models were also built and the process was repeated at four quantile points in the distribution representing the 25th, 50th, 75th and 90th percentiles (a total of 16 models). The coefficient estimates, standard error, and p-value were then used to evaluate each of the models. 

\section{Results and Findings}
 \begin{figure}[h!]
\centering
\includegraphics[width=1.1\textwidth]{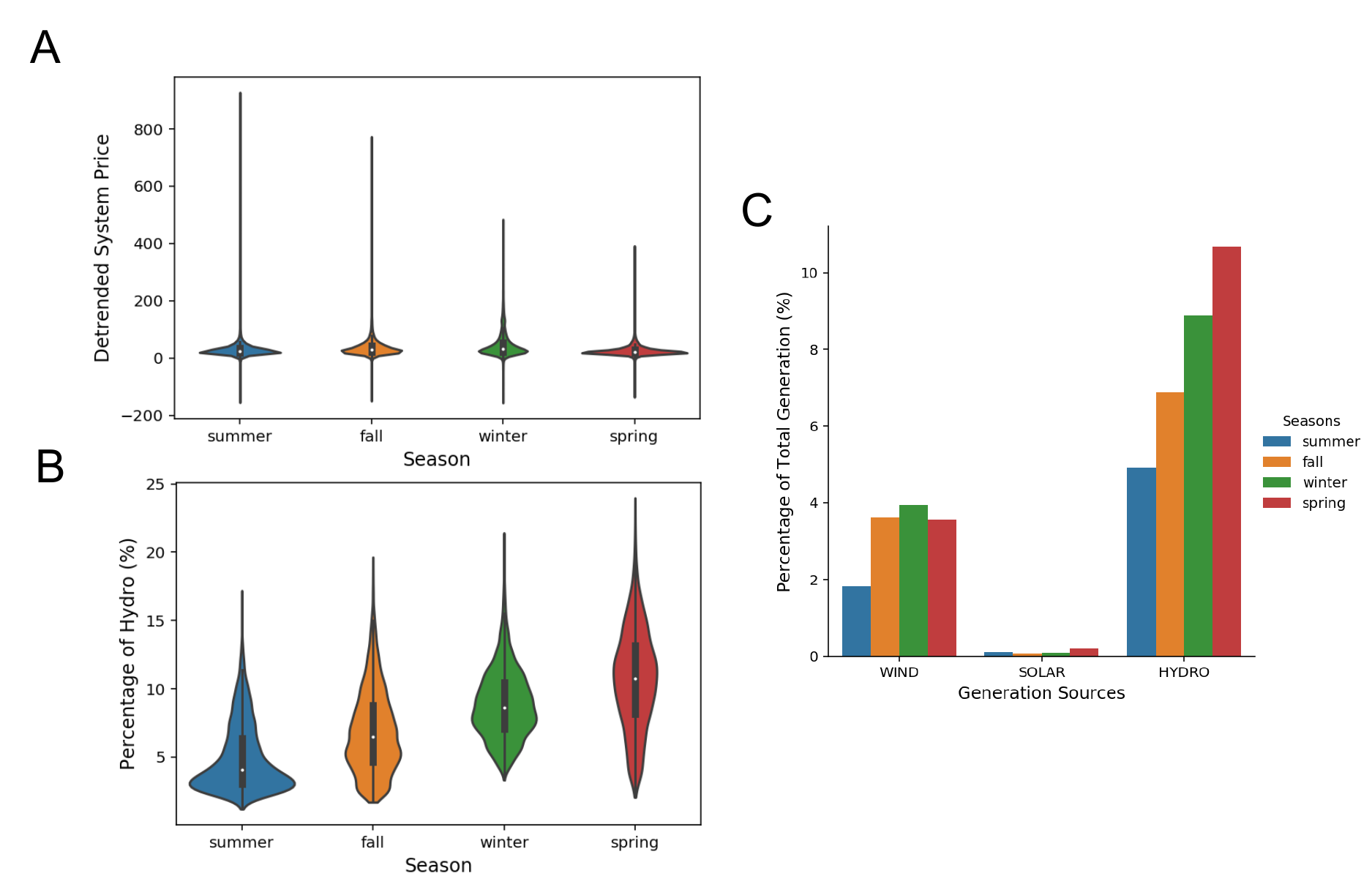}
\caption{A. Violin plots of season and detrended system price (trimmed to inner 99.99\% of price). B. Violin plots of season and percentage of hydropower in the electricity generation mix (trimmed to inner 99.99\% of price). C. Bar chart showing seasonal percentage of generation from hydropower, wind, and solar resources. We see some seasonal effects where there is a reduction in the spread of the detrended system price as the fraction of hydropower in the energy mix increases. } 
\label{fig:violin}
\end{figure}

Initial data visualizations (Fig. \ref{fig:violin}) were made to graphically understand the relationship between the percentage of hydropower present in the generation mix and the detrended
electricity price across the seasons of the year: summer, fall, winter and spring. It is important to note that while the adjustment model accounts and excludes deterministic trends in electricity price data, resulting in Fig. \ref{fig:violin}A showing no variation in the mean value across the seasons, the hydropower generation data contains seasonal patterns, as seen in Fig. \ref{fig:violin}B. Also, there are observed residual patterns in the detrended system price that is left after deterministic seasonal adjustment are corrected for, such as the variation in the range of the detrended system price for different seasons.

Summer is observed to have a large range of detrended electricity price and a low percentage of electricity coming from hydropower. In contrast, spring is observed to have a smaller range of detrended electricity prices while having a higher percentage of electricity coming from hydropower. We can observe a gradual reduction in the maximum values of detrended system price as hydropower penetration increases. Though solar and wind also have seasonal patterns (see Fig. \ref{fig:solar plot} and Fig. \ref{fig:wind plot}), their penetration is less than hydropower, as seen in Fig. \ref{fig:violin}C. 

\begin{table}[h!]
    \centering
\begin{adjustbox}{width=1\textwidth}
\begin{tabular}{llrrr}
\toprule
Model & Coefficients & Estimate & Std. Error& P-value\\
\midrule
 & $\beta_0$ & 40.07 & 0.38& 0.00\\

 \multirow{-3}{*}{\raggedright\arraybackslash $\beta_0$ + $\beta_1*hydro$}& $\beta_1$ & -0.71 & 0.04 & 0.00\\


\cmidrule{1-5}
 & $\beta_0$ & 40.26 & 0.38& 0.00\\
 
 & $\beta_1$ & -0.67 & 0.04& 0.00\\

 \multirow{-4}{*}{\raggedright\arraybackslash $\beta_0 + \beta_1*hydro +  \beta_2*solar$}& $\beta_2$ & -4.85 & 0.56& 0.00\\





\cmidrule{1-5}
 & $\beta_0$ & 41.89 & 0.40& 0.00\\
 
 & $\beta_1$ & -0.60 & 0.04& 0.00\\

 \multirow{-4}{*}{\raggedright\arraybackslash $\beta_0 + \beta_1*hydro +  \beta_2*wind$}& $\beta_2$ & -0.85 & 0.06& 0.00\\
 
 \cmidrule{1-5}
 & $\beta_0$ & 41.98 & 0.40& 0.00\\
 
 & $\beta_1$ & -0.56 & 0.04& 0.00\\
 
  & $\beta_2$ & -0.82 & 0.06& 0.00\\

 \multirow{-5}{*}{\raggedright\arraybackslash $\beta_0 + \beta_1*hydro +  \beta_2*wind + \beta_3*solar$}& $\beta_3$ & -4.33 & 0.56& 0.00\\
\bottomrule
\end{tabular}
\end{adjustbox}

    \caption{Average Effects: MLR with detrended system price as the response variable}
    \label{tab:mlr_price}
 \end{table}   
 
 \begin{table}[h!]
    \centering
\begin{adjustbox}{width=1\textwidth}
\begin{tabular}{llrrr}
\toprule
Model & Coefficients & Estimate & Std. Error & P-value\\
\midrule
 & $\beta_0$ & 6.24 & 0.15 & 0.00\\

 \multirow{-3}{*}{\raggedright\arraybackslash $\beta_0 + \beta_1*hydro$}& $\beta_1$ & -0.03 & 0.02 & 0.00\\


\cmidrule{1-5}
 & $\beta_0$ & 6.21 & 0.16 & 0.00\\
 
 & $\beta_1$ & -0.04 & 0.02 & 0.02\\

 \multirow{-4}{*}{\raggedright\arraybackslash $\beta_0 + \beta_1*hydro +  \beta_2*solar$}& $\beta_2$ & -1.60 & 0.24 & 0.00\\





\cmidrule{1-5}
 & $\beta_0$ & 6.00 & 0.16 & 0.00\\
 
 & $\beta_1$ & -0.04 & 0.02 & 0.03\\

 \multirow{-4}{*}{\raggedright\arraybackslash $\beta_0 + \beta_1*hydro +  \beta_2*wind$}& $\beta_2$ & -0.13 & 0.03 & 0.00\\
 
 \cmidrule{1-5}
 & $\beta_0$ & 6.24 & 0.17 & 0.00\\
 
 & $\beta_1$ & -0.03 & 0.02 & 0.153\\
 
  & $\beta_2$ & -0.11 & 0.03 & 0.00\\

 \multirow{-5}{*}{\raggedright\arraybackslash $\beta_0 + \beta_1*hydro +  \beta_2*wind + \beta_3*solar$}& $\beta_3$ & -1.53 & 0.24 & 0.00\\
\bottomrule
\end{tabular}
\end{adjustbox}

    \caption{Average Effects: MLR with detrended system price volatility as the response variable}
    \label{tab:mlr_vol}
\end{table}
 
\subsection{Mean Effects of Hydropower on Detrended System Price and Detrended Price Volatility}
The results of the Multiple Linear Regression (MLR) for hydropower and detrended system price using only the fraction of hydropower in the energy generation mix as the single feature in the model show that hydropower has a statistically significant negative coefficient, indicating that hydropower decreases the mean detrended system electricity price (Table \ref{tab:mlr_price}).
Additional models of hydropower coupled with either wind or solar (hydro + wind and hydro + solar) also show significant negative coefficients for hydropower, as well as solar and wind, indicating that all three contribute to a decrease in detrended system electricity price. However, we observe a weaker effect for hydropower for reducing the mean detrended system price compared to other forms of renewables, as seen by the smaller magnitude of the hydropower coefficient. When all three energy variables are included in the MLR model (hydro + wind + solar), we observe that a 1\% increase in the percentage of hydropower in the energy mix contributes to a reduction in the detrended system electricity price by \$0.56/MWh.

The MLR result for hydropower and detrended price volatility using only the fraction of hydropower in the energy generation mix as the single feature in the model shows that hydropower has a statistically significant negative coefficient, thus decreases the detrending price volatility. The models that combine hydropower with either solar or wind (hydro + wind and hydro + solar) both show that hydropower is a statistically significant contributor alongside wind and solar to the reduction in electricity price volatility.  
However, when all three energy variable are included in the MLR model, wind and solar have statistically significant negative coefficients, while the hydropower coefficient is negative but not statistically significant. Across all models in Table \ref{tab:mlr_price}, the reduction in the detrended system price volatility is less from hydropower when compared to solar and wind, as seen by the smaller magnitude in the coefficient.

\subsection{Quantile Effects of Hydropower on Price and Price Volatility}
Table \ref{tab:qr_price} shows the QR model result for detrended electricity price at the 90th percentile ($\tau$ = 0.9). We see a significant negative slope when using only the fraction of hydropower in the energy generation mix as the single feature in the model. For the combination of hydropower with either solar or wind (hydro + solar or hydro + wind), we observe a negative coefficient for all three energy sources. One may notice that the magnitude of the coefficients for solar are substantially larger than hydropower or wind. This may be an artifact of the low presence of utility-scale solar energy in ISONE, as seen in Fig. \ref{fig:violin}C. When all three energy variable are included in the QR model (hydro + wind + solar), all three energy sources still show statistically significant negative effects (p-value less than 0.05) on decreasing the 90th percentile of the detrended system price. At the 90th percentile of the detrended system price, using the joint QR model (hydro + wind + solar), we observe that a 1\% increase in the percentage of hydropower in the energy mix contributes to a reduction in the detrended system electricity price by \$1.36/MWh. Figure 4 is a plot of the coefficient of the joint QR model (hydro + wind + solar) on the detrended electricity price across quantiles. In general, 
above the 70th percentile, hydropower tends to have a larger influence on decreasing electricity price. 

Table \ref{tab:qr_vol} shows the QR at the 90th percentile of each of the four models tested (hydro, hydro + solar, hydro + wind, and hydro + wind + solar) against detrended system price volatility. When using only the fraction of hydropower in the energy generation mix as the single feature in the model, hydropower has a statistically significant negative coefficient. The hydro + solar model shows a statistically significant negative coefficient for both hydropower and solar penetration on price volatility. The hydro + wind model, however, shows a significant negative slope for hydropower penetration but an insignificant negative coefficient for wind. For the joint model (hydro + wind + solar), we see a statistically significant negative coefficient for both hydropower and solar but an insignificant positive coefficient for wind penetration. At the 90th percentile, using the joint QR model (hydro + wind + solar), we observe that a 1\% increase in the percentage of hydropower in the energy mix contributes to a reduction in the detrended system electricity price volatility by \$0.07/MWh. Once again the magnitude of the coefficients for solar may be an artifact of the low presence of utility-scale solar energy in ISONE, as seen in Fig. \ref{fig:violin}C. Figure \ref{fig:qr_vol} shows the joint hydro + wind + solar model in relation to detrended system price volatility across all quantiles. Above the 70th percentile, we observe a large negative slope in the quantile coefficient plot for hydropower and solar, while the wind coefficient tends to increase beyond the 70th percentile. We also observe that the quantile coefficient decreases below the mean values (solid red line) at extreme quantiles for both hydropower and solar. For wind however, there is a positive increase in the coefficient above the mean line as the quantile increases indicating a contribution to increasing volatility at higher quantiles. In general, we observe that the relation at higher quantiles is not only significant but also shows a stronger effect for hydropower to reduce detrended system price and price volatility as seen by the increased magnitude of the coefficient. At higher quantiles, hydropower tends to have a strong contribution to reducing volatility, which may provide a stabilizing effect for volatile wind.
\begin{table}[h!]
  \begin{center}
    \label{tab:table1}
    \begin{adjustbox}{width=1\textwidth}
    \begin{tabular}{l l r r r}
    \hline
      Model & Coefficients & Estimate & Std. Error & P-value\\
      \hline
      \multirow{2}{*}{$\beta_0+\beta_1*hydro$} & $\beta_0$ & 74.47 & 0.88 & 0.00\\
      & $\beta_1$ & -1.61 & 0.07 & 0.00\\ 
      \hline
      \multirow{3}{*}{$\beta_0+\beta_1*hydro+\beta_2*solar$} & $\beta_0$ & 74.21 & 0.87 & 0.00\\
      & $\beta_1$ & -1.39 & 0.08 & 0.00\\ 
      & $\beta_2$ & -14.03 & 0.39 & 0.00\\
      \hline
      \multirow{3}{*}{$\beta_0+\beta_1*hydro+\beta_2*wind$} & $\beta_0$ & 78.59 & 0.88 & 0.00\\
      & $\beta_1$ & -1.55 & 0.07 & 0.00\\ 
      & $\beta_2$ & -1.35 & 0.09 & 0.00\\
      \hline
      \multirow{4}{*}{$\beta_0+\beta_1*hydro+\beta_2*wind+\beta_3*solar$} & $\beta_0$ & 77.30 & 0.98 & 0.00\\
      & $\beta_1$ & -1.36 & 0.09 & 0.00\\ 
      & $\beta_2$ & -0.97 & 0.11 & 0.00\\
      & $\beta_3$ & -12.78 & 0.43 & 0.00\\
      \hline
    \end{tabular}
    \end{adjustbox}
  \end{center}
  \caption{Quantile Effects: QR with detrended price  as the response variable ($\tau=0.9$)}
\label{tab:qr_price}
\end{table}

\begin{table}[h!]
  \begin{center}
    \label{tab:table1}
    \begin{adjustbox}{width=1\textwidth}
    \begin{tabular}{l l r r r}
    \hline
      Model & Coefficients & Estimate & Std. Error & P-value\\
      \hline
      \multirow{2}{*}{$\beta_0+\beta_1*hydro$} & $\beta_0$ & 12.21 & 0.25 & 0.00\\
      & $\beta_1$ & -0.12 & 0.03 & 0.00\\ 
      \hline
      \multirow{3}{*}{$\beta_0+\beta_1*hydro+\beta_2*solar$} & $\beta_0$ & 12.14 & 0.22 & 0.00\\
      & $\beta_1$ & -0.06 & 0.03 & 0.02\\ 
      & $\beta_2$ & -3.09 & 0.09 & 0.00\\
      \hline
      \multirow{3}{*}{$\beta_0+\beta_1*hydro+\beta_2*wind$} & $\beta_0$ & 12.29 & 0.26 & 0.00\\
      & $\beta_1$ & -0.11 & 0.03 & 0.00\\ 
      & $\beta_2$ & -0.03 & 0.04 & 0.41\\
      \hline
      \multirow{4}{*}{$\beta_0+\beta_1*hydro+\beta_2*wind+\beta_3*solar$} & $\beta_0$ & 12.10 & 0.23 & 0.00\\
      & $\beta_1$ & -0.07 & 0.03 & 0.01\\ 
      & $\beta_2$ & 0.03 & 0.03 & 0.29\\
      & $\beta_3$ & -3.12 & 0.11 & 0.00\\
      \hline
    \end{tabular}
    \end{adjustbox}
    
  \end{center}
  \caption{Quantile Effects: QR with detrended price volatility as the response variable ($\tau=0.9$)}
\label{tab:qr_vol}
\end{table}
\section{Discussion}
Increase in the percentage of hydropower in the energy generation mix is found to be correlated with a decrease in detrended system electricity price both at the mean and upper quantiles. The contributing effect of hydropower in the presence of other renewables (wind and solar) also  results in a net negative reduction in detrended electricity price. Hydropower is a contributor to the merit order effect \cite{senfub2008} since hydropower plants have low marginal cost and  can supply cheaper electricity in the wholesale market, thus lowering the system price. This is consistent with studies on the average (mean effects) relationship between hydropower and price \cite{Wen2022, Pereira2017}. While our results confirm the average reduction in system electricity price with increased generation from hydropower, the reduction is greater when considering extreme electricity prices. On average (mean effects), hydropower contributes less to reducing the mean system electricity price relative to solar or wind. At the mean, the marginal contribution of hydropower to a reduction in the mean system electricity price is 32\% less than the contribution from wind. However, at the 90th percentile,  hydropower contributes 41\% more to the reduction in detrended system electricity price than wind. 

Increase in the percentage of hydropower in the energy generation mix is found to be correlated with a decrease in detrended system electricity price volatility at the mean and upper quantiles both independently and in the presence of wind and solar. However, our results indicate that hydropower's contribution to reduced volatility is larger and more significant at extremes (periods of high price volatility). When all three energy sources are considered, the marginal contribution of hydropower to a reduction in the mean system electricity price volatility is 73\% less than the contribution from wind and statistically insignificant. However, at the 90th percentile, hydropower contributes to a decrease in detrended system price volatility, while wind increases the volatility. The magnitude of the contribution to the detrended price volatility reduction from hydropower is more than twice the contribution from wind to the price volatility increase. The contribution of hydropower to decreasing volatility may be explained by the storage characteristics of hydropower where water is pumped up the reservoir at low prices (low demand period) and released at higher prices (higher demand period), thus producing a smoothing effect. 
 \begin{figure}[h!]
\centering
\includegraphics[width=12.5cm]{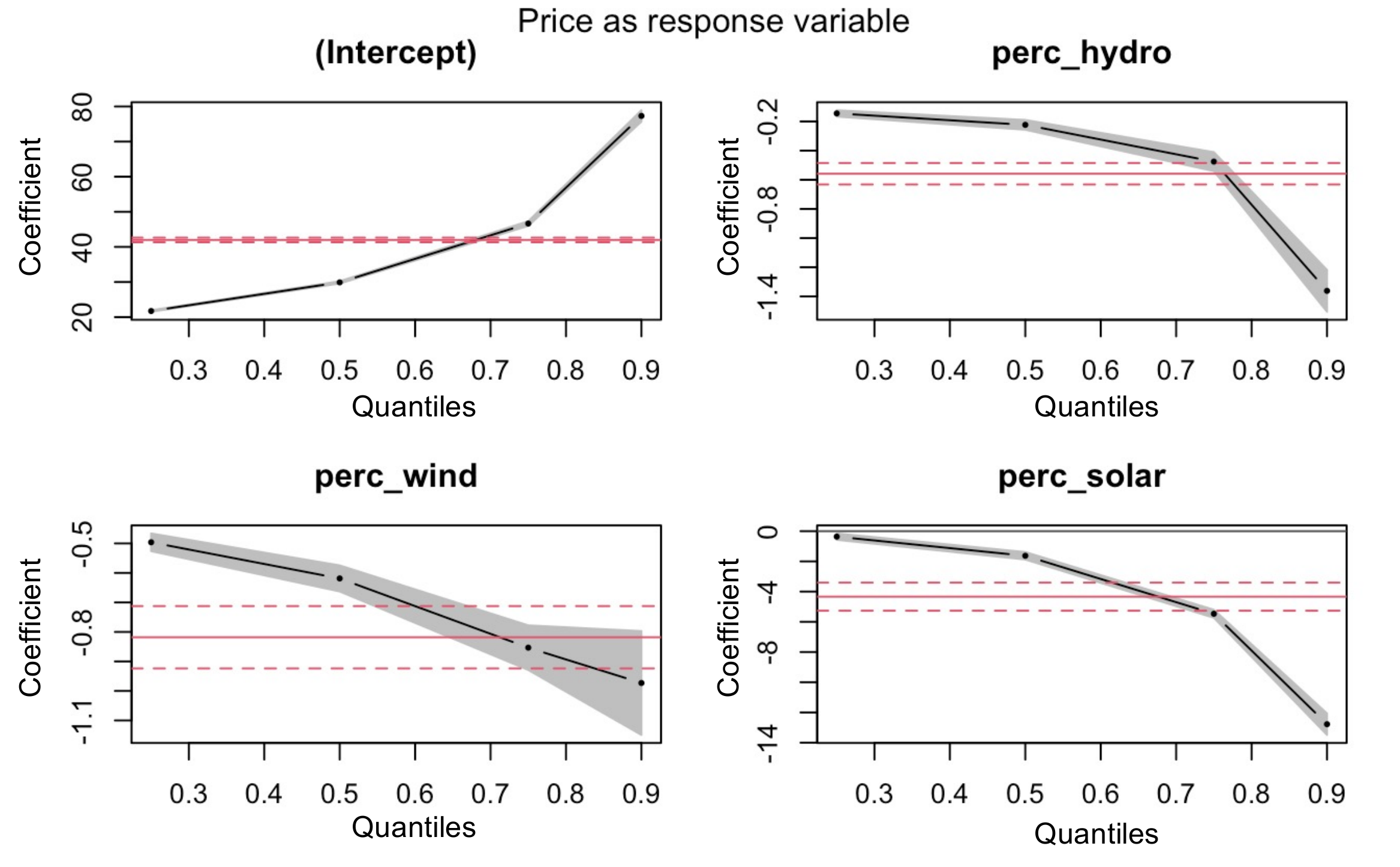}
\caption{Quantile Coefficient Plot. This shows the coefficient for the joint quantile model (hydro + wind + solar) on price. The solid red line represents the least square regression estimates while the red dashed line represents the 95\% confidence interval. The black line represent the quantile regression coefficient across each of 25th, 50th, 75th and 90th percentile (represented with the black dots) while the gray shaded region indicates the corresponding error margin.} 
\label{fig:qr_price}
\end{figure}

 \begin{figure}[h!]
\centering
\includegraphics[width=12.5cm]{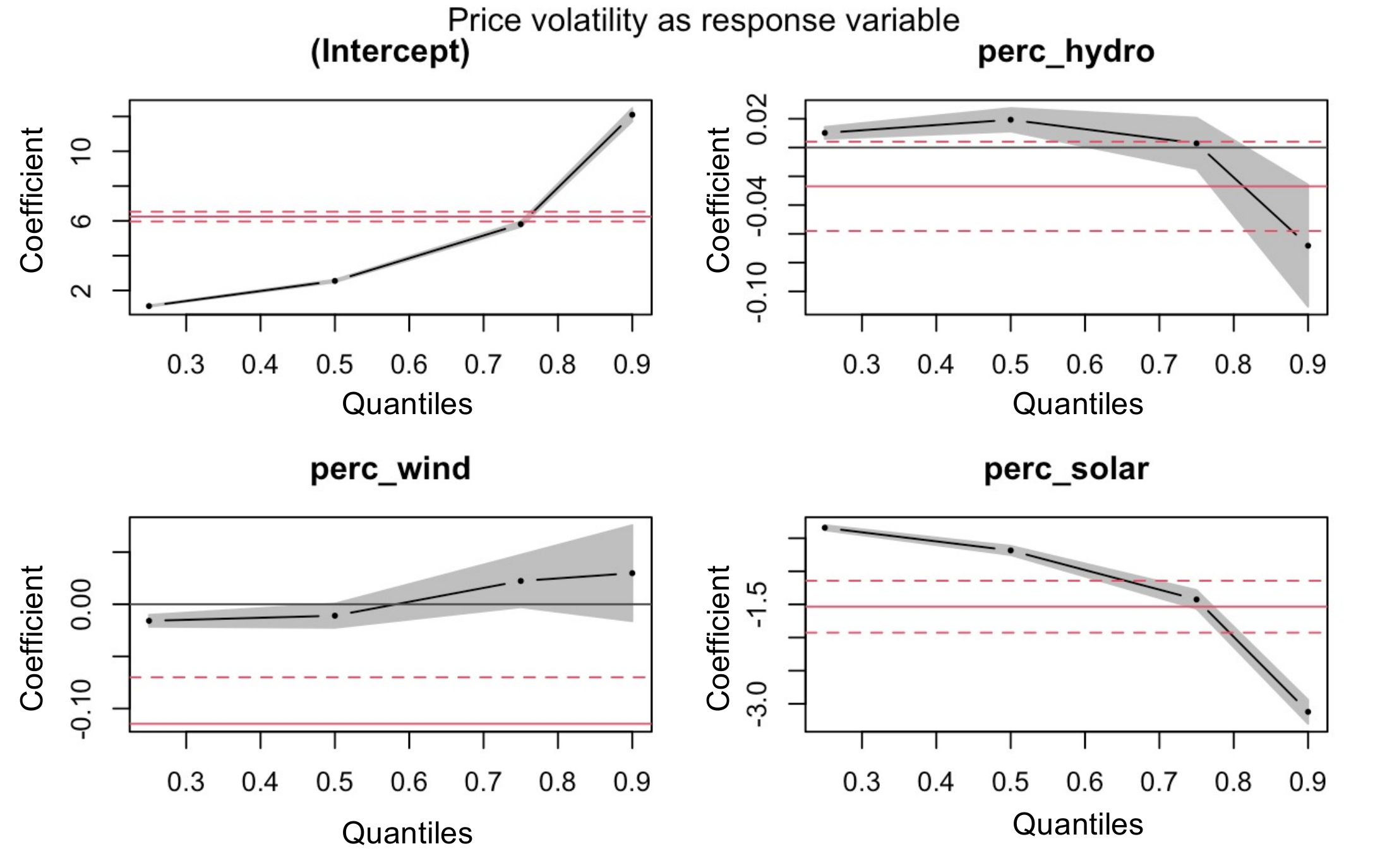}
\caption{Quantile Coefficient Plot. This shows the coefficient for the joint quantile model (hydro + wind + solar) on price volatility. The solid red line represents the least square regression estimates while the red dashed line represents the 95\% confidence interval. The black line represent the quantile regression coefficient across eaach of 25th., 50th, 75th and 90th percentile (represented with the black dots) while the gray shaded region indicates the corresponding error margin.} 
\label{fig:qr_vol}
\end{figure}



\section{Conclusion}
In this study, we conducted an analysis of the role of hydropower on electricity price and price volatility. The analysis is based on the detrended system electricity price data after adjusting for seasonal, weekend, and hourly effects. We found that the adjustment model used in this study can help to expose actual trends that may be dampened by behavioral and temporal characteristics of price, and therefore improve accurate interpretations of the role of hydropower on price and price volatility.  We then built models to explore the mean and quantile effects of hydropower in the presence of solar and wind. The conditional models were especially useful for observing patterns at extremes of price and price volatility. Findings on the average (mean effects) reduction in price as the percentage of hydropower increases in the energy generation mix are consistent with literature \cite{Wen2022, Pereira2017} and the merit order effect \cite{senfub2008}. Additionally, we find that hydropower contributes to a greater reduction in system price at higher quantiles. For price volatility, hydropower provides a small and sometimes insignificant reduction to the mean price volatility. However, at the extreme quantiles, hydropower contributes a significant decrease in price volatility while offsetting the effect from wind which tends to increase volatility.

Our finding that the effect of hydropower on electricity price and price volatility is  different at the mean and upper quantiles is important for making accurate and actionable conclusions. This is particularly useful for designing policy incentives and plans for new and existing hydropower plants. The storage capabilities of hydropower allow it to act as a shock absorber for price volatility, reducing extreme energy prices and price volatility. While this study has evaluated the effect of hydropower on price and price volatility in ISONE, it is not meant as a universal conclusion on this relationship. The analysis is dependent on factors that vary regionally, such as the amount of hydropower present, the composition of the market's energy portfolio, the market and incentive structure, and seasonal and temporal patterns of electricity demand. Also, direct interpretation of the results should be limited to the bounds of the data used in this study. For example, utility-scale solar has a much lower installed capacity in ISONE relative to hydropower and wind and care must be taken to avoid extrapolation when interpreting its coefficients. The study results are valid within the studied temporal range (2014 - 2020) and the hydropower, wind, and solar penetration levels.  Given global and regional ambitious renewable energy targets, future penetration levels are likely to increase to match these targets and, therefore, the future relationships may differ from the levels observed in this study.
Additionally, further research should incorporate quantitative data on hydropower storage, such as reservoir capacity, percentage full, and environmental constraints, in order to quantify the impact of hydropower as a storage resource on electricity price and price volatility. While this study has explored linear quantiles, future studies should explore non-linear quantile analysis, as these might expose non-linear effects in the data. 



\section*{Acknowledgments}
\textbf{Funding:} \\
NSF Harnessing the Data Revolution (HDR) program, ``Collaborative Research: Predictive Risk Investigation SysteM (PRISM) for Multi-layer Dynamic Interconnection Analysis" (\#1940176, 1940223, 2023755). U.S. DOE, ``Wave Energy Technology Assessment for Optimal Grid Integration and Blue Economy Advancement" (DE-EE9443).\\
\textbf{Author Contributions:}\\
Conceptualization: CS, DAS, LW, MGS\\
Methodology: OOO, DAS, CS, LW\\
Investigation: OOO, SS, LW, YH \\
Visualization: OOO\\
Data Curation: OOO\\
Formal Analysis: OOO, SS, LW, YH\\
Funding acquisition: CS, DAS, LW, MGS\\
Project administration: OOO, CS, DAS, LW, MGS\\
Supervision: CS, DAS, LW, MGS\\
Writing – original draft: OOO\\
Writing – review \& editing: OOO, SS, YH, KL, CS, DAS, LW, MGS\\
\textbf{Competing Interests:}\\
CS and KL have financial interests in HydroSapient, Inc., a company which could potentially benefit from the results of this research. This interest has been reviewed by the University in accordance with its Individual Conflict of Interest policy, for the purpose of maintaining the objectivity and the integrity of research at The Pennsylvania State University.\\

\clearpage
\appendix
\section{Appendix}
\subsection{Tables}

\begin{table}[h!]
  \begin{center}
    \label{tab:table1}
    \begin{adjustbox}{width=1\textwidth}
    \begin{tabular}{l l r r r}
    \hline
      Model & Coefficients & Estimate & Std. Error & P-value\\
      \hline
      \multirow{2}{*}{$\beta_0+\beta_1*hydro$} & $\beta_0$ & 20.78 & 0.11 & 0.00\\
      & $\beta_1$ & -0.23 & 0.01 & 0.00\\ 
      \hline
      \multirow{3}{*}{$\beta_0+\beta_1*hydro+\beta_2*solar$} & $\beta_0$ & 20.85 & 0.11 & 0.00\\
      & $\beta_1$ & -0.23 & 0.01 & 0.00\\ 
      & $\beta_2$ & -0.77 & 0.13 & 0.00\\
      \hline
      \multirow{3}{*}{$\beta_0+\beta_1*hydro+\beta_2*wind$} & $\beta_0$ & 21.74 & 0.12 & 0.00\\
      & $\beta_1$ & -0.15 & 0.01 & 0.00\\ 
      & $\beta_2$ & -0.50 & 0.02 & 0.00\\
      \hline
      \multirow{4}{*}{$\beta_0+\beta_1*hydro+\beta_2*wind+\beta_3*solar$} & $\beta_0$ & 21.74 & 0.12 & 0.00\\
      & $\beta_1$ & -0.14 & 0.01 & 0.00\\ 
      & $\beta_2$ & -0.50 & 0.02 & 0.00\\
      & $\beta_3$ & -0.36 & 0.13 & 0.00\\
      \hline
    \end{tabular}
    \end{adjustbox}
  \end{center}
  \caption{Quantile Effects: QR with detrended price as the response variable ($\tau=0.25$)}
\label{table:1}
\end{table}

\begin{table}[h!]
  \begin{center}
    \label{tab:table1}
    \begin{adjustbox}{width=1\textwidth}
    \begin{tabular}{l l r r r}
    \hline
      Model & Coefficients & Estimate & Std. Error & P-value\\
      \hline
      \multirow{2}{*}{$\beta_0+\beta_1*hydro$} & $\beta_0$ & 28.43 & 0.19 & 0.00\\
      & $\beta_1$ & -0.33 & 0.02 & 0.00\\ 
      \hline
      \multirow{3}{*}{$\beta_0+\beta_1*hydro+\beta_2*solar$} & $\beta_0$ & 28.52 & 0.19 & 0.00\\
      & $\beta_1$ & -0.30 & 0.02 & 0.00\\ 
      & $\beta_2$ & -2.00 & 0.15 & 0.00\\
      \hline
      \multirow{3}{*}{$\beta_0+\beta_1*hydro+\beta_2*wind$} & $\beta_0$ & 29.77 & 0.20 & 0.00\\
      & $\beta_1$ & -0.23 & 0.02 & 0.00\\ 
      & $\beta_2$ & -0.63 & 0.03 & 0.00\\
      \hline
      \multirow{4}{*}{$\beta_0+\beta_1*hydro+\beta_2*wind+\beta_3*solar$} & $\beta_0$ & 29.90 & 0.19 & 0.00\\
      & $\beta_1$ & -0.22 & 0.02 & 0.00\\ 
      & $\beta_2$ & -0.62 & 0.03 & 0.00\\
      & $\beta_3$ & -1.63 & 0.15 & 0.00\\
      \hline
    \end{tabular}
    \end{adjustbox}
  \end{center}
  \caption{Quantile Effects: QR with detrended price as the response variable ($\tau=0.5$)}
\label{table:1}
\end{table}

\begin{table}[h!]
  \begin{center}
    \label{tab:table1}
    \begin{adjustbox}{width=1\textwidth}
    \begin{tabular}{l l r r r}
    \hline
      Model & Coefficients & Estimate & Std. Error & P-value\\
      \hline
      \multirow{2}{*}{$\beta_0+\beta_1*hydro$} & $\beta_0$ & 44.85 & 0.37 & 0.00\\
      & $\beta_1$ & -0.68 & 0.04 & 0.00\\ 
      \hline
      \multirow{3}{*}{$\beta_0+\beta_1*hydro+\beta_2*solar$} & $\beta_0$ & 45.25 & 0.36 & 0.00\\
      & $\beta_1$ & -0.62 & 0.04 & 0.00\\ 
      & $\beta_2$ & -6.56 & 0.16 & 0.00\\
      \hline
      \multirow{3}{*}{$\beta_0+\beta_1*hydro+\beta_2*wind$} & $\beta_0$ & 46.71 & 0.41 & 0.00\\
      & $\beta_1$ & -0.54 & 0.04 & 0.00\\ 
      & $\beta_2$ & -0.96 & 0.06 & 0.00\\
      \hline
      \multirow{4}{*}{$\beta_0+\beta_1*hydro+\beta_2*wind+\beta_3*solar$} & $\beta_0$ & 46.65 & 0.40 & 0.00\\
      & $\beta_1$ & -0.48 & 0.04 & 0.00\\ 
      & $\beta_2$ & -0.85 & 0.05 & 0.00\\
      & $\beta_3$ & -5.47 & 0.16 & 0.00\\
      \hline
    \end{tabular}
    \end{adjustbox}
  \end{center}
  \caption{Quantile Effects: QR with detrended price as the response variable ($\tau=0.75$)}
\label{table:1}
\end{table}

\begin{table}[h!]
  \begin{center}
    \label{tab:table1}
    \begin{adjustbox}{width=1\textwidth}
    \begin{tabular}{l l r r r}
    \hline
      Model & Coefficients & Estimate & Std. Error & P-value\\
      \hline
      \multirow{2}{*}{$\beta_0+\beta_1*hydro$} & $\beta_0$ & 1.056 & 0.020 & 0.000\\
      & $\beta_1$ & 0.004 & 0.002 & 0.074\\ 
      \hline
      \multirow{3}{*}{$\beta_0+\beta_1*hydro+\beta_2*solar$} & $\beta_0$ & 1.079 & 0.020 & 0.000\\
      & $\beta_1$ & 0.007 & 0.002 & 0.003\\ 
      & $\beta_2$ & -0.350 & 0.024 & 0.000\\
      \hline
      \multirow{3}{*}{$\beta_0+\beta_1*hydro+\beta_2*wind$} & $\beta_0$ & 1.095 & 0.022 & 0.000\\
      & $\beta_1$ & 0.007 & 0.002 & 0.003\\ 
      & $\beta_2$ & -0.019 & 0.003 & 0.000\\
      \hline
      \multirow{4}{*}{$\beta_0+\beta_1*hydro+\beta_2*wind+\beta_3*solar$} & $\beta_0$ & 1.107 & 0.022 & 0.000\\
      & $\beta_1$ & 0.010 & 0.003 & 0.000\\ 
      & $\beta_2$ & -0.016 & 0.003 & 0.000\\
      & $\beta_3$ & -0.341 & 0.023 & 0.000\\
      \hline
    \end{tabular}
    \end{adjustbox}
  \end{center}
  \caption{Quantile Effects: QR with detrended price volatility as the response variable ($\tau=0.25$)}
\label{table:1}
\end{table}

\begin{table}[h!]
  \begin{center}
    \label{tab:table1}
    \begin{adjustbox}{width=1\textwidth}
    \begin{tabular}{l l r r r}
    
    \hline
      Model & Coefficients & Estimate & Std. Error & P-value\\
      \hline
      \multirow{2}{*}{$\beta_0+\beta_1*hydro$} & $\beta_0$ & 2.49 & 0.04 & 0.00\\
      & $\beta_1$ & 0.01 & 0.00 & 0.02\\ 
      \hline
      \multirow{3}{*}{$\beta_0+\beta_1*hydro+\beta_2*solar$} & $\beta_0$ & 2.53 & 0.04 & 0.00\\
      & $\beta_1$ & 0.02 & 0.00 & 0.00\\ 
      & $\beta_2$ & -0.69 & 0.04 & 0.00\\
      \hline
      \multirow{3}{*}{$\beta_0+\beta_1*hydro+\beta_2*wind$} & $\beta_0$ & 2.52 & 0.04 & 0.00\\
      & $\beta_1$ & 0.01 & 0.00 & 0.01\\ 
      & $\beta_2$ & -0.02 & 0.01 & 0.02\\
      \hline
      \multirow{4}{*}{$\beta_0+\beta_1*hydro+\beta_2*wind+\beta_3*solar$} & $\beta_0$ & 2.55 & 0.04 & 0.00\\
      & $\beta_1$ & 0.02 & 0.00 & 0.00\\ 
      & $\beta_2$ & -0.01 & 0.01 & 0.12\\
      & $\beta_3$ & -0.68 & 0.04 & 0.00\\
      \hline
    \end{tabular}
    \end{adjustbox}
  \end{center}
  \caption{Quantile Effects: QR with detrended price volatility as the response variable ($\tau=0.5$)}
\label{table:1}
\end{table}

\begin{table}[h!]
  \begin{center}
    \label{tab:table1}
    \begin{adjustbox}{width=1\textwidth}
    \begin{tabular}{l l r r r}
    \hline
      Model & Coefficients & Estimate & Std. Error & P-value\\
      \hline
      \multirow{2}{*}{$\beta_0+\beta_1*hydro$} & $\beta_0$ & 5.87 & 0.09 & 0.00\\
      & $\beta_1$ & -0.02 & 0.01 & 0.08\\ 
      \hline
      \multirow{3}{*}{$\beta_0+\beta_1*hydro+\beta_2*solar$} & $\beta_0$ & 5.88 & 0.09 & 0.00\\
      & $\beta_1$ & 0.00 & 0.01 & 0.70\\ 
      & $\beta_2$ & -1.38 & 0.10 & 0.00\\
      \hline
      \multirow{3}{*}{$\beta_0+\beta_1*hydro+\beta_2*wind$} & $\beta_0$ & 5.89 & 0.10 & 0.00\\
      & $\beta_1$ & -0.02 & 0.01 & 0.11\\ 
      & $\beta_2$ & -0.01 & 0.02 & 0.58\\
      \hline
      \multirow{4}{*}{$\beta_0+\beta_1*hydro+\beta_2*wind+\beta_3*solar$} & $\beta_0$ & 5.81 & 0.10 & 0.00\\
      & $\beta_1$ & 0.00 & 0.01 & 0.79\\ 
      & $\beta_2$ & 0.02 & 0.02 & 0.14\\
      & $\beta_3$ & -1.42 & 0.09 & 0.00\\
      \hline
    \end{tabular}
    \end{adjustbox}
  \end{center}
  \caption{Quantile Effects: QR with detrended price volatility as the response variable ($\tau=0.75$)}
\label{table:1}
\end{table}
\clearpage
\subsection{Figures}
\begin{figure}[h!]
\centering
\includegraphics[width=10cm]{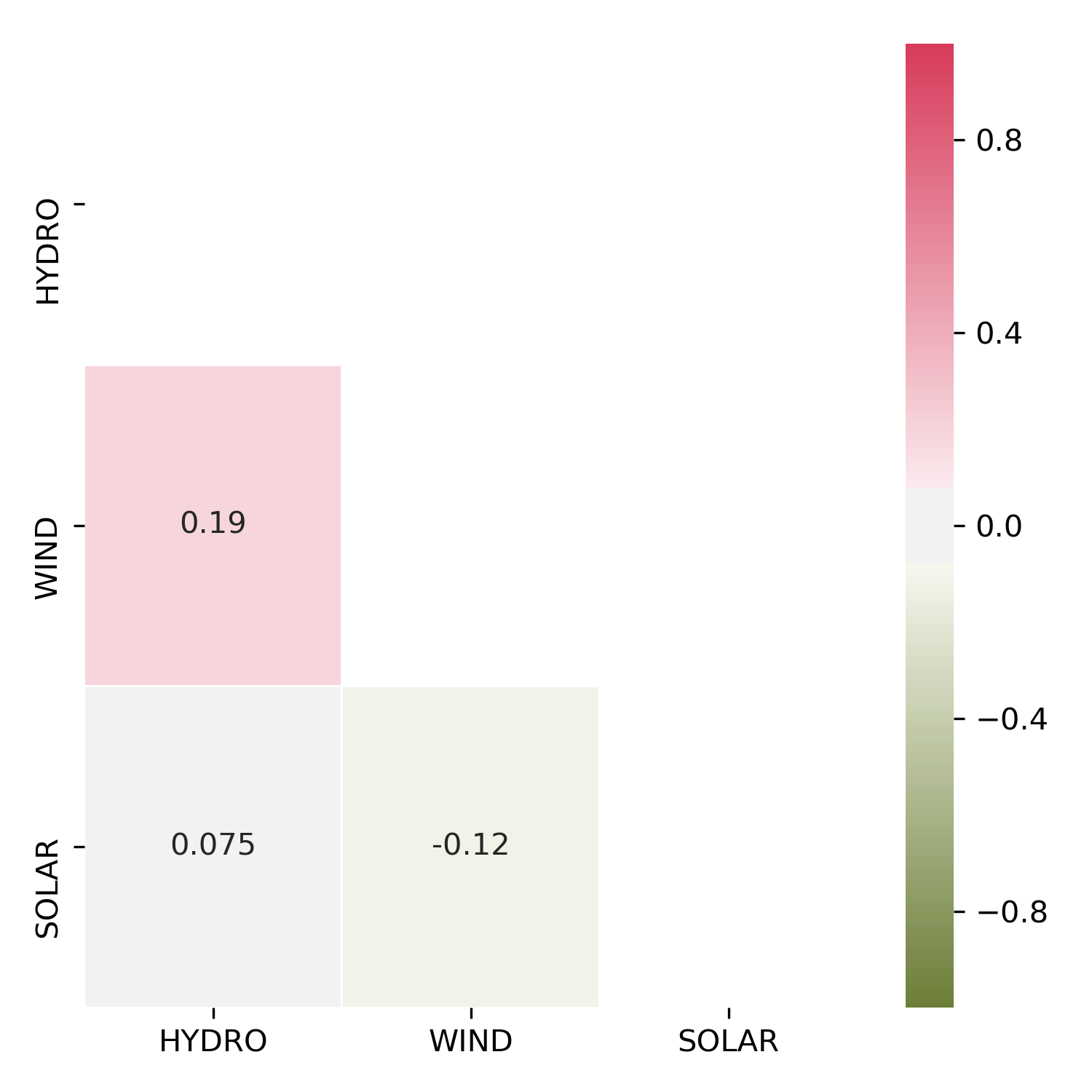}
\caption{Correlation Plot.  The correlation plot was used to diagnose possible multicollinearity issues in the MLR model. There is minimal correlation between solar, wind, and hydropower generation.} 
\label{fig:correlation plot}
\end{figure}

\begin{figure}[h!]
\centering
\includegraphics[width=12cm]{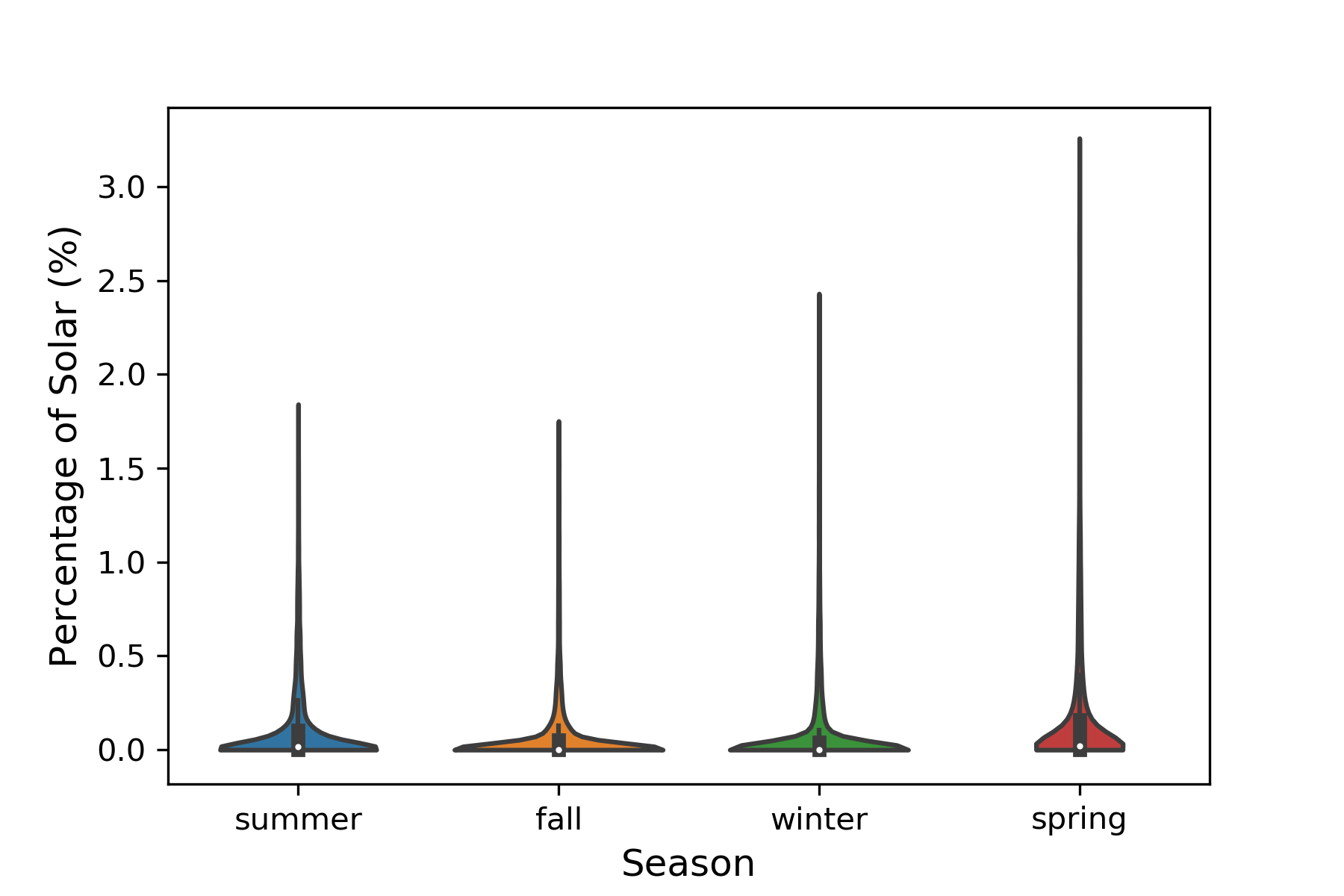}
\caption{Violin plot of solar penetration (trimmed to inner 99\% of detrended system price)} 
\label{fig:solar plot}
\end{figure}

\begin{figure}[h!]
\centering
\includegraphics[width=12cm]{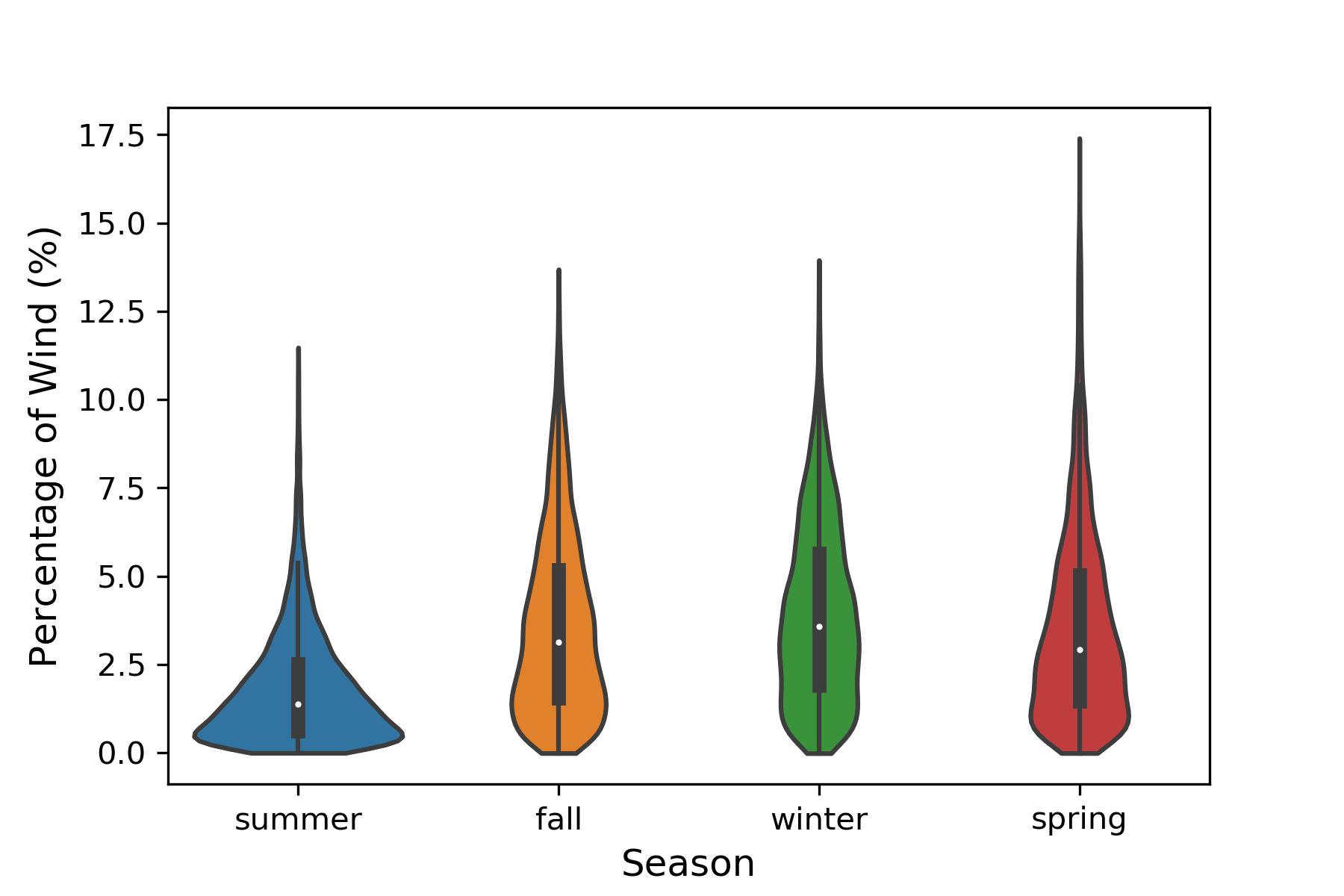}
\caption{Violin plot of wind penetration (trimmed to inner 99\% of detrended system price)} 
\label{fig:wind plot}
\end{figure}
\clearpage
\bibliographystyle{ieeetr}
\bibliography{ref}

\begin{thebibliography}{10}

\bibitem{Rehman2015}
S.~Rehman, L.~M. Al-Hadhrami, and M.~M. Alam, ``Pumped hydro energy storage
  system: A technological review,'' {\em Renewable and Sustainable Energy
  Reviews}, vol.~44, pp.~586--598, 4 2015.

\bibitem{ISONE2021}
ISONE, ``{ISO New England Web Services API v1.1}.''
  \url{https://webservices.iso-ne.com/docs/v1.1/}, 2021.
\newblock Accessed: 2021-01-10.

\bibitem{Buonocore2019}
J.~J. Buonocore, E.~J. Hughes, D.~R. Michanowicz, J.~Heo, J.~G. Allen, and
  A.~Williams, ``Climate and health benefits of increasing renewable energy
  deployment in the united states,'' {\em Environmental Research Letters},
  vol.~14, p.~114010, 10 2019.

\bibitem{frass2009renewable}
C.~Fr{\"a}ss-Ehrfeld, {\em Renewable energy sources: a chance to combat climate
  change}, vol.~1.
\newblock Kluwer Law International BV, 2009.

\bibitem{useiarenew2021}
{U.S. Energy Information Administration (EIA)}, ``Renewables became the
  second-most prevalent u.s. electricity source in 2020,'' 2021.

\bibitem{EIA2021}
EIA, ``What is u.s. electricity generation by energy source?.''
  \url{https://www.eia.gov/tools/faqs/faq.php?id=427&t=3}, 2021.
\newblock Accessed: 2021-11-08.

\bibitem{Foley2015}
A.~M. Foley, P.~G. Leahy, K.~Li, E.~J. McKeogh, and A.~P. Morrison, ``A
  long-term analysis of pumped hydro storage to firm wind power,'' {\em Applied
  Energy}, vol.~137, pp.~638--648, 1 2015.

\bibitem{Ardizzon2014}
G.~Ardizzon, G.~Cavazzini, and G.~Pavesi, ``A new generation of small hydro and
  pumped-hydro power plants: Advances and future challenges,'' {\em Renewable
  and Sustainable Energy Reviews}, vol.~31, pp.~746--761, 3 2014.

\bibitem{Kaldellis2001}
J.~K. Kaldellis and K.~A. Kavadias, ``Optimal wind-hydro solution for aegean
  sea islands' electricity-demand fulfilment,'' {\em Applied Energy}, vol.~70,
  pp.~333--354, 12 2001.

\bibitem{Bakos2002}
G.~C. Bakos, ``Feasibility study of a hybrid wind/hydro power-system for
  low-cost electricity production,'' {\em Applied Energy}, vol.~72,
  pp.~599--608, 7 2002.

\bibitem{Kapsali2010}
M.~Kapsali and J.~K. Kaldellis, ``Combining hydro and variable wind power
  generation by means of pumped-storage under economically viable terms,'' {\em
  Applied Energy}, vol.~87, pp.~3475--3485, 11 2010.

\bibitem{Kapsali2012}
M.~Kapsali, J.~S. Anagnostopoulos, and J.~K. Kaldellis, ``Wind powered
  pumped-hydro storage systems for remote islands: A complete sensitivity
  analysis based on economic perspectives,'' {\em Applied Energy}, vol.~99,
  pp.~430--444, 11 2012.

\bibitem{Glasnovic2009}
Z.~Glasnovic and J.~Margeta, ``The features of sustainable solar hydroelectric
  power plant,'' {\em Renewable Energy}, vol.~34, pp.~1742--1751, 7 2009.

\bibitem{Margeta2010}
J.~Margeta and Z.~Glasnovic, ``Feasibility of the green energy production by
  hybrid solar + hydro power system in europe and similar climate areas,''
  {\em Renewable and Sustainable Energy Reviews}, vol.~14, pp.~1580--1590, 8
  2010.

\bibitem{Margeta2012}
J.~Margeta and Z.~Glasnovic, ``Theoretical settings of photovoltaic-hydro
  energy system for sustainable energy production,'' {\em Solar Energy},
  vol.~86, pp.~972--982, 3 2012.

\bibitem{Zhao2012}
J.~Zhao, K.~Graves, C.~Wang, G.~Liao, and C.~P. Yeh, ``A hybrid electric/hydro
  storage solution for standalone photovoltaic applications in remote areas,''
  {\em IEEE Power and Energy Society General Meeting}, 2012.

\bibitem{Javanbakht2013}
P.~Javanbakht, S.~Mohagheghi, and M.~G. Simoes, ``Transient performance
  analysis of a small-scale pv-phs power plant fed by a svpwm drive applied for
  a distribution system,'' {\em 2013 IEEE Energy Conversion Congress and
  Exposition, ECCE 2013}, pp.~4532--4539, 2013.

\bibitem{Ma2015}
T.~Ma, H.~Yang, L.~Lu, and J.~Peng, ``Pumped storage-based standalone
  photovoltaic power generation system: Modeling and techno-economic
  optimization,'' {\em Applied Energy}, vol.~137, pp.~649--659, 1 2015.

\bibitem{Zhang2019}
H.~Zhang, Z.~Lu, W.~Hu, Y.~Wang, L.~Dong, and J.~Zhang, ``Coordinated optimal
  operation of hydro–wind–solar integrated systems,'' {\em Applied Energy},
  vol.~242, pp.~883--896, 5 2019.

\bibitem{Wen2022}
L.~Wen, K.~Suomalainen, B.~Sharp, M.~Yi, and M.~S. Sheng, ``Impact of
  wind-hydro dynamics on electricity price: A seasonal spatial econometric
  analysis,'' {\em Energy}, vol.~238, p.~122076, 1 2022.

\bibitem{Woo2011}
C.~K. Woo, I.~Horowitz, J.~Moore, and A.~Pacheco, ``The impact of wind
  generation on the electricity spot-market price level and variance: The texas
  experience,'' {\em Energy Policy}, vol.~39, pp.~3939--3944, 7 2011.

\bibitem{Ketterer2014}
J.~C. Ketterer, ``The impact of wind power generation on the electricity price
  in germany,'' {\em Energy Economics}, vol.~44, pp.~270--280, 7 2014.

\bibitem{Suomalainen2015}
K.~Suomalainen, G.~Pritchard, B.~Sharp, Z.~Yuan, and G.~Zakeri, ``Correlation
  analysis on wind and hydro resources with electricity demand and prices in
  new zealand,'' {\em Applied Energy}, vol.~137, pp.~445--462, 1 2015.

\bibitem{Pereira2017}
J.~P. Pereira, V.~Pesquita, and P.~M. Rodrigues, ``The effect of hydro and wind
  generation on the mean and volatility of electricity prices in spain,'' {\em
  International Conference on the European Energy Market, EEM}, 7 2017.

\bibitem{Somani2021}
A.~Somani, N.~Voisin, R.~Tipireddy, S.~Turner, T.~D. Veselka, Q.~Ploussard,
  V.~Koritarov, T.~M. Mosier, M.~Mohanpurkar, M.~R. Ingram, S.~Signore,
  B.~Hadjerioua, B.~T. Smith, P.~W. O'connor, and R.~Shan, ``Hydropower value
  study: Current status and future opportunities,'' 2021.

\bibitem{ISONEabout2021}
ISONE, ``About us.'' \url{https://www.iso-ne.com/about}, 2021.
\newblock Accessed: 2021-11-08.

\bibitem{ISONEresource2021}
ISONE, ``Resource mix.''
  \url{https://www.iso-ne.com/about/key-stats/resource-mix/}, 2021.
\newblock Accessed: 2021-11-08.

\bibitem{isoneoutlook2020}
ISONE, ``{2020 Regional Electricity Outlook}.''
  \url{https://www.iso-ne.com/static-assets/documents/2020/02/2020\_reo.pdf},
  2020.
\newblock Accessed: 2021-11-08.

\bibitem{isonelmp}
ISONE, ``Locational marginal pricing.''
  \url{https://www.iso-ne.com/participate/support/faq/lmp}, 2021.
\newblock Accessed: 2021-11-08.

\bibitem{Yoo2014}
W.~Yoo, R.~Mayberry, S.~Bae, K.~Singh, Q.~P. He, J.~W. Lillard, and Jr., ``A
  study of effects of multicollinearity in the multivariable analysis,'' {\em
  International journal of applied science and technology}, vol.~4, p.~9, 10
  2014.

\bibitem{Bernard2015}
C.~Bernard and C.~Czado, ``Conditional quantiles and tail dependence,'' {\em
  Journal of Multivariate Analysis}, vol.~138, pp.~104--126, 6 2015.

\bibitem{lane20212}
J.~Lane, ``{Robust Quantile Regression Using L2E},'' {\em Rice University},
  2012.

\bibitem{senfub2008}
F.~Sensfuß, M.~Ragwitz, and M.~Genoese, ``The merit-order effect: A detailed
  analysis of the price effect of renewable electricity generation on spot
  market prices in germany,'' {\em Energy Policy}, vol.~36, pp.~3086--3094, 8
  2008.

\end{thebibliography}
\end{document}